\documentclass{aa}  

\usepackage{xcolor}
\usepackage{yfonts}

\usepackage[normalem]{ulem}

\usepackage{lipsum}
\usepackage{yfonts}

\usepackage{graphicx}
\usepackage{txfonts}
\begin{document}

   \title{Finding flares in Kepler and TESS data with recurrent deep neural networks}

   \author{Kriszti\'an Vida
          \inst{1}
          \and
          Attila B\'odi\inst{1,3}
          \and
          Tam\'as Szklen\'ar\inst{1}
          \and
          B\'alint Seli\inst{1,2}
         } 
\institute{
Konkoly Observatory, Research Centre for Astronomy and Earth Sciences, E\"otv\"os Lor\'and Research Network (ELKH), Konkoly Thege Mikl\'os \'ut 15-17, H-1121 Budapest, Hungary\\
\email{vidakris@konkoly.hu}
\and
E\"otv\"os Lor\'and University, P\'azm\'any P\'eter s\'et\'any 1/A,  Budapest, Hungary
\and
MTA CSFK Lendület Near-Field Cosmology Research Group
}

   \date{Received September 15, 1996; accepted March 16, 1997}

 
 \abstract{
 Stellar flares are an important aspect of magnetic activity -- both for stellar evolution and circumstellar habitability viewpoints -- but automatically and accurately finding them is still a challenge to researchers in the Big Data era of astronomy. We present an experiment to detect flares in space-borne photometric data using deep neural networks. Using a set of artificial data and real photometric data we trained a set of neural networks, and found that the best performing architectures were the recurrent neural networks (RNNs) using Long Short-Term Memory (LSTM) layers. The best trained network detected flares over {$5\sigma$ with $\gtrsim80$\% recall and precision} and was also capable to distinguish typical false signals (e.g. maxima of RR Lyr stars) from real flares. Testing the network trained on Kepler data on TESS light curves showed that the  neural net is able to generalize and find flares -- with similar effectiveness -- in completely new data having previously unseen sampling and characteristics.  }


   \maketitle
%

\section{Introduction}

Stellar flares are the result of magnetic field line reconnection -- they appear as sudden brightening of the stellar atmosphere as a portion of magnetic energy is released. 
Flares are most common in late-type M-dwarfs \citep{2011AJ....141...50W}, but they appear in a wide range of main-sequence stars, and even in some evolved giant stars \citep{flaringgiants}.
These eruptions are interesting not only from a stellar astronomy viewpoint: they can also influence their surroundings, and have serious effects on the atmosphere evolution of their orbiting planets \citep{2007AsBio...7..167K, 2008SSRv..139..437Y, trappist1}.

Recent and upcoming photometric space missions, like Kepler \citep{Kepler}, TESS \citep{TESS} and PLATO \citep{PLATO} provide an unprecedented opportunity for astronomers to study these events. But this opportunity comes with a new challenge -- while earlier observations of a single object could be handled relatively easily by manual analysis, investigating photometric data of hundreds, or even thousands of targets manually is not feasible. 

There are several existing approaches to detect these transient events in space automatically. These are often based on smoothing the light curve and detecting the outliers by some criteria \citep{appaloosa, 2016MNRAS.463.1844S}, or other way of outlier detection 
(e.g. the RANSAC method, see \citealt{flatwrm}).
A known issue of these methods is that they are prone  to  misidentify other astrophysical phenomena as flares (e.g., maxima of  KIC 1572802, an RR Lyr\ae\, star), 
and they also need user attention.

These issues can be possibly remedied by deep learning methods -- a neural network, once trained by appropriate training data, does not need any user input or parametrization, and can operate autonomously on new data, and hopefully, the "astrophysical noise" (like RR Lyr peaks marked as flares) can be also filtered out. 
While the training of such networks is time consuming -- a single iteration can have a runtime of several hours even on recent GPUs -- analysis of new data is relatively fast: for example, face recognition can be done on the fly in video streams on mobile phones.

Using simple dense neural networks would be impractical for this case, since that approach would yield too many variables, making the algorithm too slow. For time series data Convolutional Neural Networks \citep[CNNs,][]{CNN_basic_article} are a much better choice, as these are much faster or better adapted for such data. 
Convolutional networks take advantage of the hierarchical pattern in data and assemble more complex patterns using smaller and simpler patterns. 
Recently, \cite{CNN-code} presented a method to find flares using CNNs.

Another approach is using Recurrent Neural Networks (RNNs): here, connections between nodes form a directed graph along a temporal sequence. This allows them to "remember" their previous state, making them suitable for a wide range of problems from improved machine translation \citep{lstm-translation} to anomaly detection in ECG time signals \citep{ecg}. This can give them the power not just simply detect outlying points, but also compare the analyzed data to previously seen light curve parts, just as a human would do, making it a powerful tool for scientists.

In this paper we present our first results of a flare detection method based on recurrent neural networks. In Sect. \ref{sect:input}. we show the data preparation steps, in Sect. \ref{sect:architecture}. the design and in Sect. \ref{sect:evaluation}. the evaluation of tested networks. In Sect. \ref{sect:kfold}-\ref{sect:validation}. we present the validation steps, while in Sec. \ref{sect:performance}. we discuss the performance of our network. We dedicate Sect. \ref{sect:failures} to present out dead ends and finally, we summarize our results in Sect. \ref{sect:summary}.

\section{Input data}
\label{sect:input}

Our training data consisted of two sets: artificial data and real Kepler observations. For the artificial data we generated light curves of spotted stars using 
PyMacula \citep{macula}\footnote{\url{https://pypi.org/project/pymacula/}}. 
{The artificial light curves were generated with different rotation periods. The periods of the light curves were chosen from a lognormal distribution with a mean of $\log5$ days, and the lowest period was set to 0.4 days. This distribution was selected to represent the most active, fast-rotating stars that are more likely to show flares. To these light curves a randomly selected number of spots (between 2--12) were added with spot evolution enabled.}
Then, we added noise with normal distribution with a noise level of 
{$8\times10^{-3}-4\times10^{-5}$ (in normalized flux units), that correspond to typical short-cadence noise levels for 7--16 magnitude stars%
\footnote{
    \url{https://nexsci.caltech.edu/workshop/2012/keplergo/CalibrationSN.shtml}
}.
}
These light curves were injected with flares using the analytical flare model of \cite{appaloosa}, 
using FWHM values between 0.5--1 hour with uniform distribution and amplitudes with lognormal distribution with a mean of $10^{-5}$ and a standard deviation of 1. The number of injected flare per star varied between 5 and 200.
The code \texttt{ocelot-gym} used to create spotted light curves injected with flares for the training is available on GitHub.\footnote{\url{https://github.com/vidakris/flatwrm2}} 

Real Kepler observations with flares were selected from the targets of \cite{sauces} with additional targets as 'astrophysical noise' (RR Lyr, Cepheids, flaring and non-flaring eclipsing binaries, systems with exoplanets, irregular variable stars). The selected short cadence light curves were first flagged with the FLATW'RM code \citep{flatwrm}, and these flagged data were checked twice manually and re-flagged if needed. This set consisted of 214 light curves, that were separated to 'active' (107) and 'quiet' (207) sets for the first steps of the model selection.
For the training we used only short-cadence data, since interpolating the long-cadence light curves would add too much information.

We found that using both time and intensity data as input the candidate neural nets did not converge, therefore we interpolated our data to 1-min cadence, and used only the flux values as a one-dimensional input vector. This, on one hand, 
helps some issues with the time axis (e.g. normalization), and also makes the analysis faster. 
On the other hand, this limits the usage of the trained network: while a data with 5-min cadence might be still handled fine, interpolating a long-cadence observation with 30-minute sampling adds too much information to the data, and also renders the analysis unnecessarily slow, therefore, for such data a different training will be needed.
Gaps in the data were interpolated and filled with random noise with normal distribution -- with characteristics of the light curve -- to avoid jumps in the light curve that can mimic flares. 

Flux data were also standardized, after dropping NaN non-numeric values, the observed flux was first 
divided by its truncated mean, then centered around zero by subtracting 1, finally we divided by the truncated peak-to-peak variation. 
The truncated mean and peak-to-peak variation were calculated after dropping flux values above the 99.9th percentile of the data (leaving out large flares). Experience shows that machine learning algorithms perform best with standardized data having a variation more or less 1.

The prepared light curves were concatenated and used as a single one-dimensional input vector for the training. Training can be made faster with training multiple batches simultaneously (the batch number is mainly limited by GPU memory). However, creating batches with time series is not obvious, as the consecutive data parts are not independent from each other. Therefore -- as opposed to the default time series generator in Keras -- we created the batches by separating the data into $b$ consecutive data segments, that feeds these batches after each other into the network for training. 
Using too large batch number might hinder the convergence of the network (especially if the length of the resulting batches are in the magnitude of the events we are looking for), but we found a batch number of $b=2048$ a good compromise. 
With this setup, the runtime of a single epoch was 12--17 minutes on an NVidia GeForce RTX 2080 Ti graphics card.

For the analysis each light curve point was associated with a flag to mark if it is part of a flare or not.
This flag was determined based on 64-point windows starting from the given data point: the labels were calculated as the average of the input flags rounded to 0 or 1. In the case of the final data points,  where the number of the remaining points is less than 64, the original data points were mirrored in order to avoid losing any data.
Our experiments with different window sizes and different ways of calculating the labels, e.g. including exceptions for short events, or aligning the windows differently,  all yielded poorer results (see also Sect. \ref{sect:failures}).

{After selecting the most successful network candidate we returned to the input: we used its output to re-examine the training data. We checked the output for possible missed events that was not marked as flares in the original set, or light curve regions we marked incorrectly as flares. The network was then re-trained with this refined training set. These corrections improved the performance (recall/accuracy) up to 10\% in the case of smaller events.}

\section{Network architecture}
\label{sect:architecture}

\begin{table}
\caption{Parameters of the tested network architectures. }
\label{tab:hparams} 
\centering
    
\begin{tabular}{l|cc}
\hline
\hline
Parameter & Tested values & Chosen value  \\
\hline
Kernel           &   [GRU, LSTM] & LSTM \\
Kernel units     &   [128, 256]  & 128 \\
Number of recurrent layers      &   [2, 3]      & 3 \\
Dense layers &   [0, 1]      & 0 \\
Dropout rate     &   [0.2, 0.5]  & 0.2 \\
\hline
Optional dense layers & \multicolumn{2}{c}{512, sigmoid}  \\
(units, activation) & \multicolumn{2}{c}{64, sigmoid}  \\
\hline
Output dense layer & \multicolumn{2}{c}{1, sigmoid}  \\
(units, activation)&&\\
\hline

Loss function & \multicolumn{2}{c}{Binary crossentropy}  \\
Optimizer        &   \multicolumn{2}{c}{NAdam (learning rate=0.001)}  \\
Metrics         & \multicolumn{2}{c}{Accuracy, Recall} \\

    \end{tabular}

\end{table}
\begin{figure}
    \centering
    \includegraphics[width=0.45\textwidth]{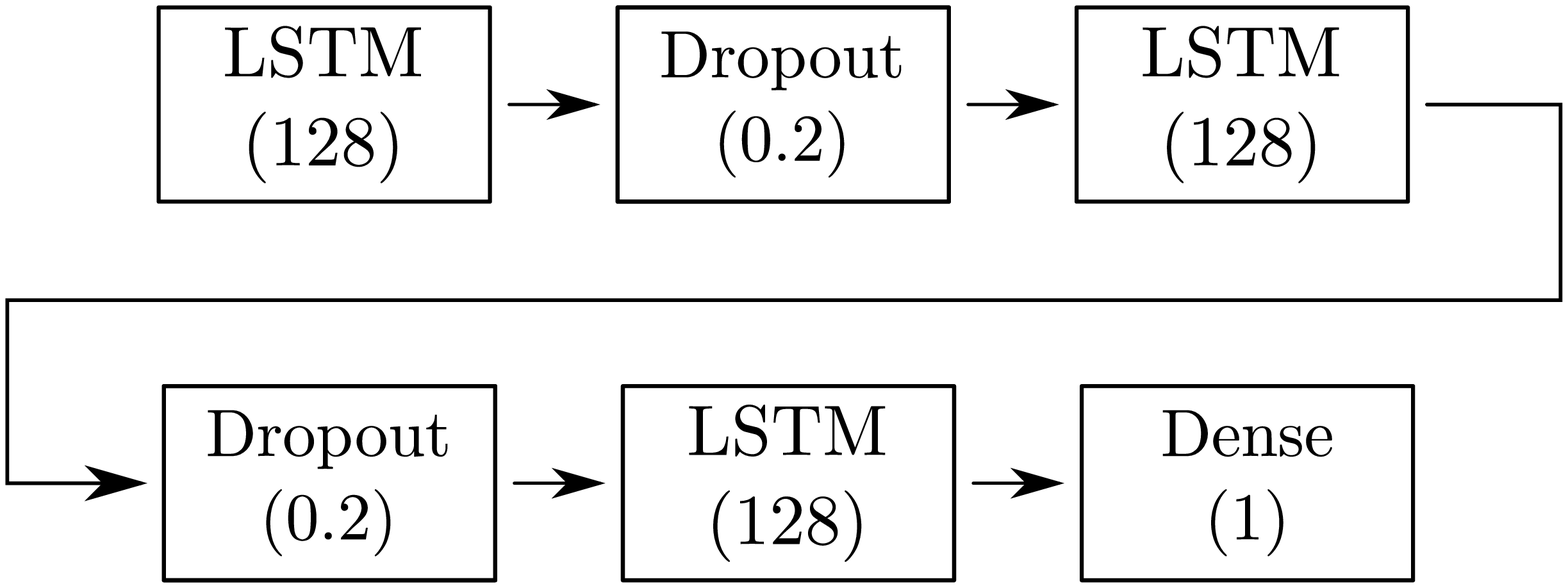}
    \caption{The final neural network architecture consisted of three LSTM layers with 128 units each, and a dropout layer between them with dropout rate of 0.2. The output was generated by a one-unit dense layer with sigmoid activation function. }
    \label{fig:architecture}
\end{figure}

\begin{figure*}
    \centering
    \includegraphics[width=0.5\textwidth]{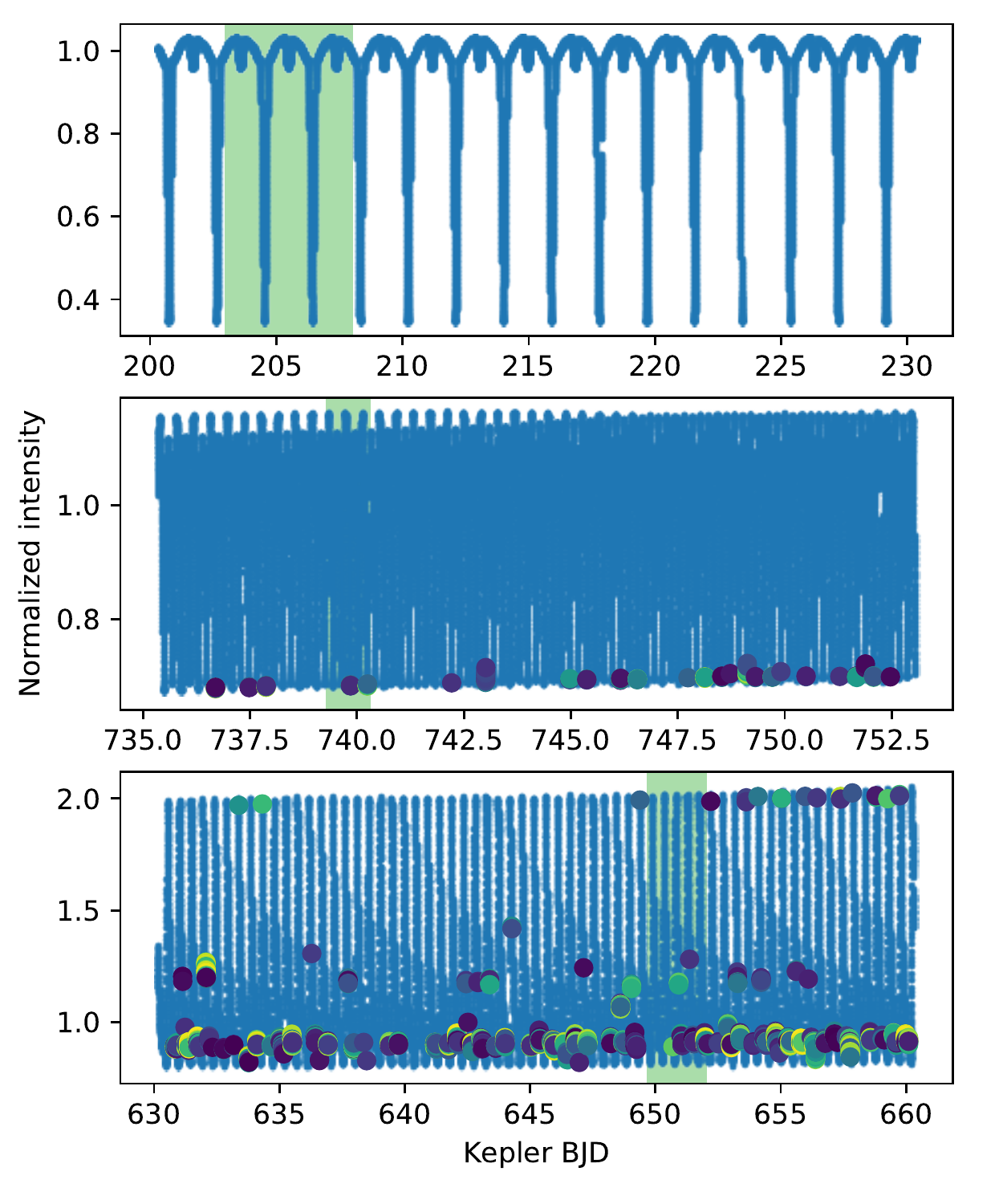}%
    \includegraphics[width=0.35\textwidth]{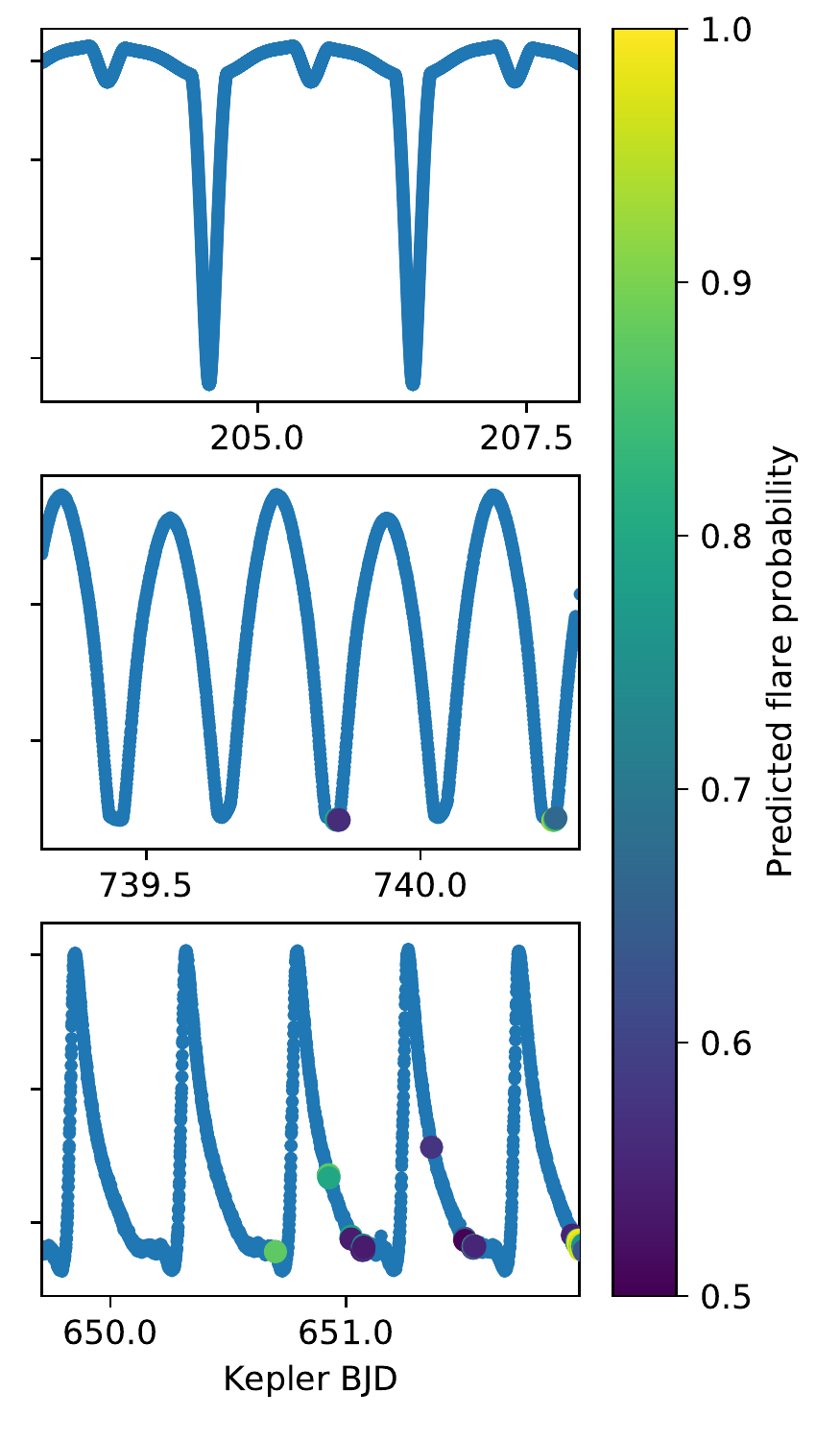}
    \caption{Examples from the set used as 'astrophysical noise'. Larger points indicate points predicted as possible flares by the trained network.
    From top to bottom: Algol type-, $\beta$ Lyrae type eclipsing binary, RR Lyrae. 
    The few-point false negative detections make post-processing and validation a necessary, but also a relatively easy task. The false positive detections in the plots will be cleaned by median filtering before the validation step. 
    }
    \label{fig:noisedata}
\end{figure*}

The neural network was built using Keras\footnote{\url{keras.io}}, 
a deep learning API built on Tensorflow\footnote{\url{https://www.tensorflow.org}}.
We experimented with a large number of network architectures (see Sect. \ref{sect:failures}), but our final network candidates consisted of sequential models with two or three GRU/LSTM layers with 128/256 units (Gated Recurrent Unit, \citealt{gru-cho2014learning}; Long Short-Term Memory layer, \citealt{lstm-hochreiter1997}). 
Between the recurrent layers a dropout layer \citep{dropout-srivastava2014} was added with a dropout rate of 0.2 -- utilizing such layers in a networks serves as a kind of regularization, by temporarily removing a fraction of the neurons randomly from the network.
This way our model can achieve lower generalization error, thus prevent overfitting, at a cost of somewhat increased computation time. 
After the recurrent layers a single one-unit dense layer was added with sigmoid activation to generate the final output (Fig. \ref{fig:architecture}). 
The networks were optimized using a binary crossentropy loss function and an NAdam optimizer \citep{adam-kingma2014, nadam}.
As the ratio of flaring and non-flaring data points is far from equal, weighting is important during the fit, so the model pays more attention to samples from the under-represented class -- we weighted the flaring points with their ratio to the total number of points.

The first stage of the model selection was performed using the \texttt{HParams} tool of Keras, which allowed us to compare a variety of model architectures. The options tried in this step of the hyperparameter tuning are summarized in Table \ref{tab:hparams} (see also Appendix \ref{sect:appendix} for more details). 
In this stage we trained the networks on our artificial data and the flaring Kepler data set. 
From the best performing models the first five best-performing networks were selected during hyperparameter tuning. 
Then, we added the non-flaring 'astrophysical noise' observations (see Fig. \ref{fig:noisedata}) to the training sample to find a network that not only can detect flares, but also neglect irrelevant signals. 
While with the flaring data the GRU networks performed well, adding the non-flaring data produced a large number of false positive detections.

\section{Evaluation}
\label{sect:evaluation}

After training the candidate networks, we ran a prediction for the validation data set, and compared them based on their 
accuracy (the rate of correctly classified points), 
classification error rate (rate of incorrectly classified points), 
recall (ratio of recovered flare points), 
precision (ratio of correct detection in the selected points), 
false positive rate, 
and $F_\beta$ values:
$$F_\beta = \frac{(1+\beta^2)\times \mathrm{TP}}{(1+\beta^2)\times \mathrm{TP} + \beta\times \mathrm{FN} + \mathrm{FP} }. $$
Here TP, FN, and FP stand for true positives, false negatives and false positives, respectively, and $\beta=1$. 
$F_\beta$ score gives the weighted harmonic mean of precision and recall. 
We found, that the best-performing networks were the two-layered LSTM(256) with or without additional CNN layers, and the three-layered LSTM(128) networks. We selected the latter candidate as the final choice, as its runtime was $\approx 20-40\%$ lower (see Appendix \ref{sect:appendix}).

Evaluating the performance of recurrent neural networks (RNN) is not as straightforward as in case of e.g. binary classifiers, where usually we have a set of train and test data with an input label for each item, which can be easily compared to the output label with the highest probability (for the complexity of the task see e.g. \citealt{RNNevaluate}). Particularly, in case of the flare detection using temporal information the problem comes from the lack of solid ground truth. First, we have to decide which kind of phenomenon is considered to be a flare and have to decide which points are part of that flare like event and which are not. As a possible solution we split the light curves and fit each input and predicted segment with a model individually and use the fits as "labels". The fitting procedure is described in Sect. \ref{sect:validation}. In the following, we bin the flares by their signal-to-noise ratio and relative amplitude and calculate the aforementioned metrics for each bin, and compare the resulted curves instead of single numbers. As it is nearly impossible to force the RNN to select the same points for a given flare that we flagged as input, the fitting may result in different number of flares within the same event if its structure is complex. Therefore we consider input-output flares the same if they lie within 0.02 days.

\section{K-Fold cross-validation}
\label{sect:kfold}

To check the robustness of the neural network, we performed K-Fold cross-validation on the training data. 
This method is commonly used in applied machine learning to compare models and to see how dependent the trained network is on the particular set of training data. 
The K-Fold cross-validation method is based on separating a set of test data (we used 10\% of the total data), and shuffling the remaining data into non-overlapping training and validation sets multiple times (we used 5 runs). 
The network is then trained on each shuffled set, yielding five slightly different trained networks with different starting weights and different training data. 
The result of this test is shown in Fig. \ref{fig:kfold}: above $3-5\sigma$ signal level the difference between the five models is only a few percent in both the precision and recall metrics, suggesting a robust model. 

\section{Post-processing and validation}
\label{sect:validation}
\begin{figure}
    \centering
    \includegraphics[width=0.5\textwidth]{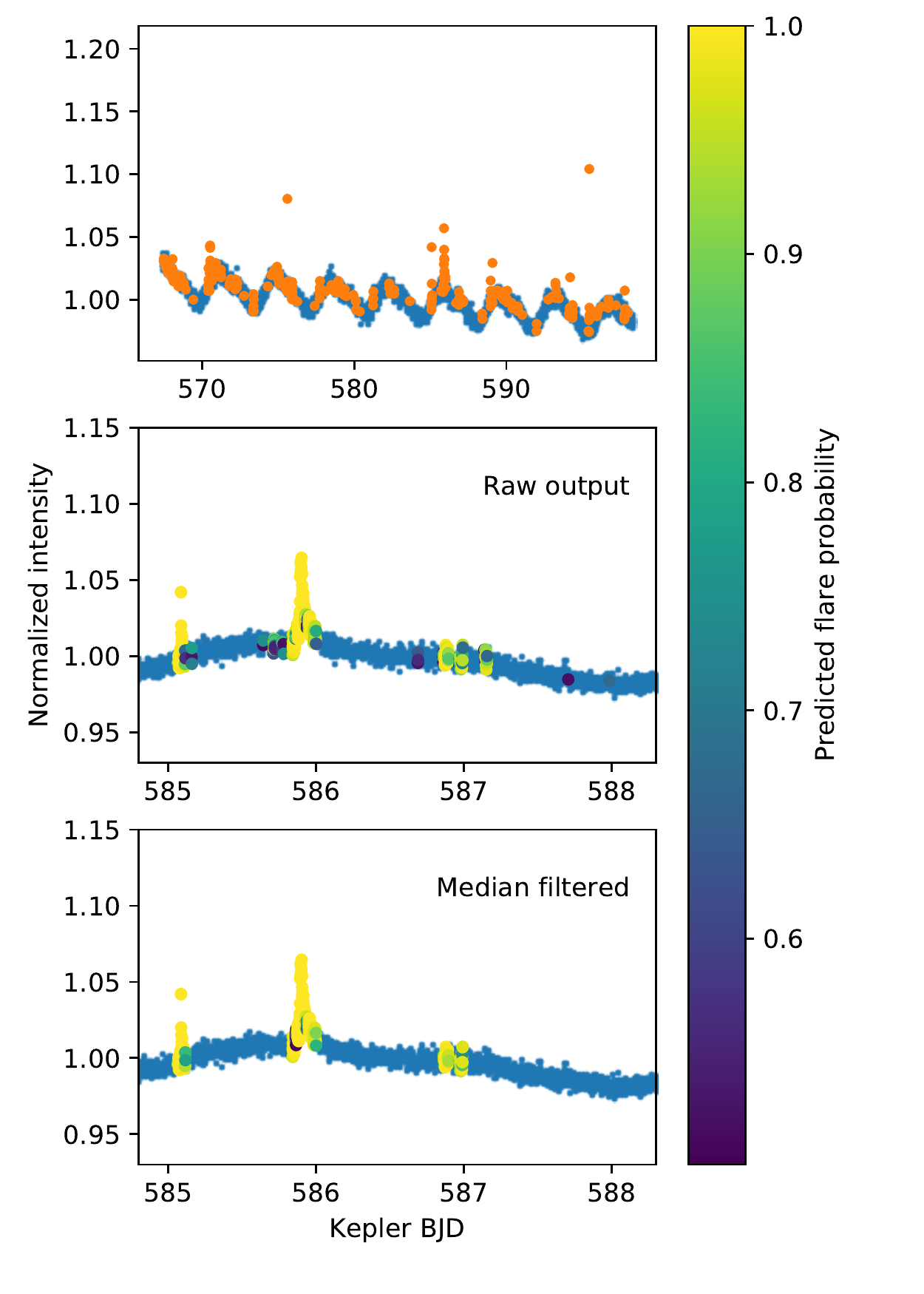}
    \caption{Example output from the validation data set (KIC 012156549). The top panel shows the input light curve (blue points)  with all the data points marked by the neural network as flares (orange points). The middle panel is zoomed in,  color code showing the raw output from the predictions where the output probability higher than 0.5. 
    There are a few points within flares that are not marked (false negatives), and some single points that are incorrectly identified as flares (false positives). 
    This can be rectified by smoothing the output probabilities by a median filter (bottom plot panel).}
    \label{fig:example-lc}
\end{figure}

\begin{figure}
    \centering
    \includegraphics[width=0.5\textwidth]{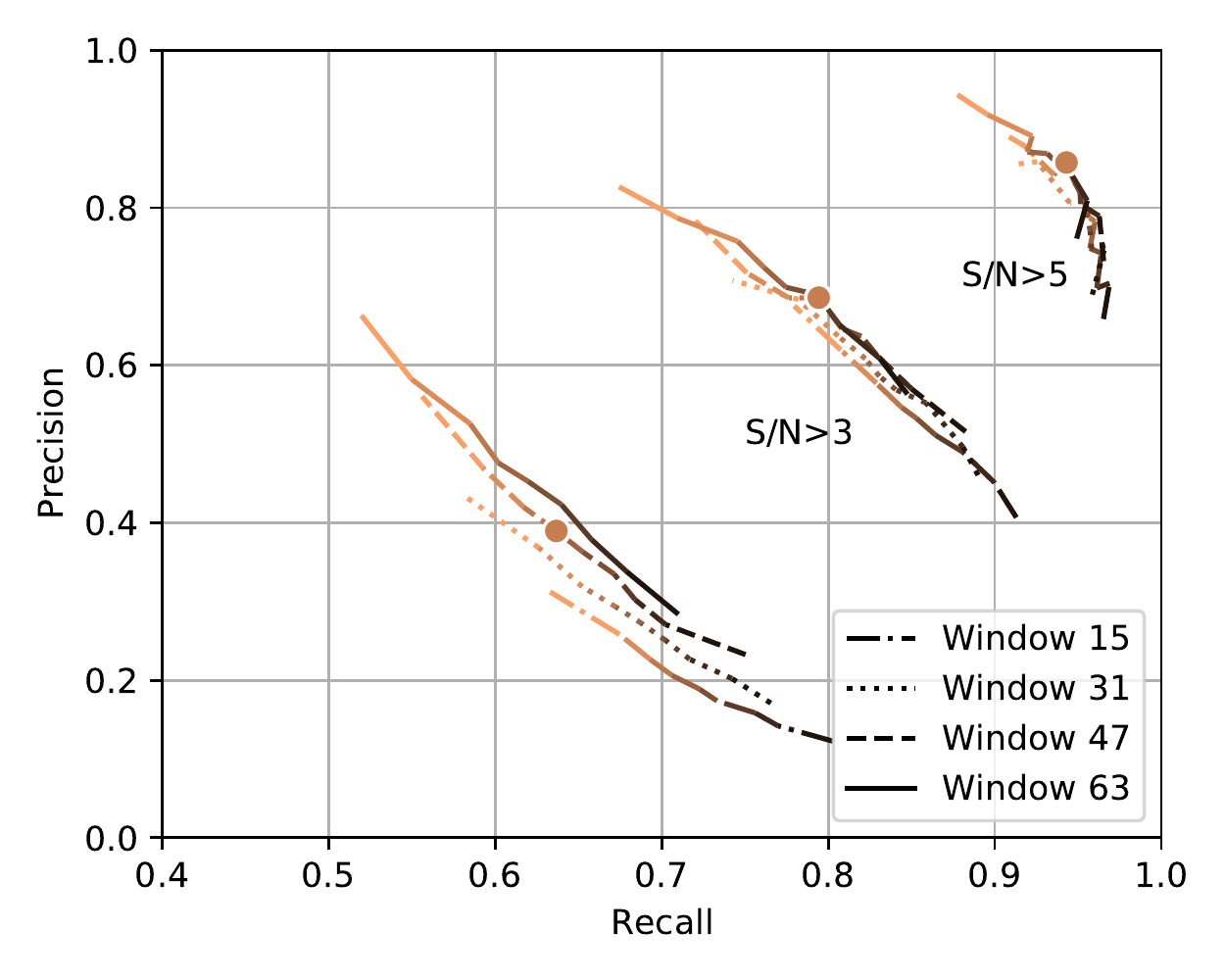}
    \caption{Precision--recall curve with median filters of different kernel sizes. The three batches show all flare events from the test data (left), and cases with flares,  in which the highest point compared to the noise in the light curve is above $3\sigma$ (middle) and $5\sigma$ (right). Color coding shows different cutoff levels above which a flux measurement is considered to be a flare. The darker the color the higher the cutoff threshold. The point marks the location of cutoff value 0.2 for 47-point-width median filtered case. See text for details. }
    \label{fig:pre-recall}
\end{figure}

\begin{figure}
    \centering
    \includegraphics[width=0.5\textwidth]{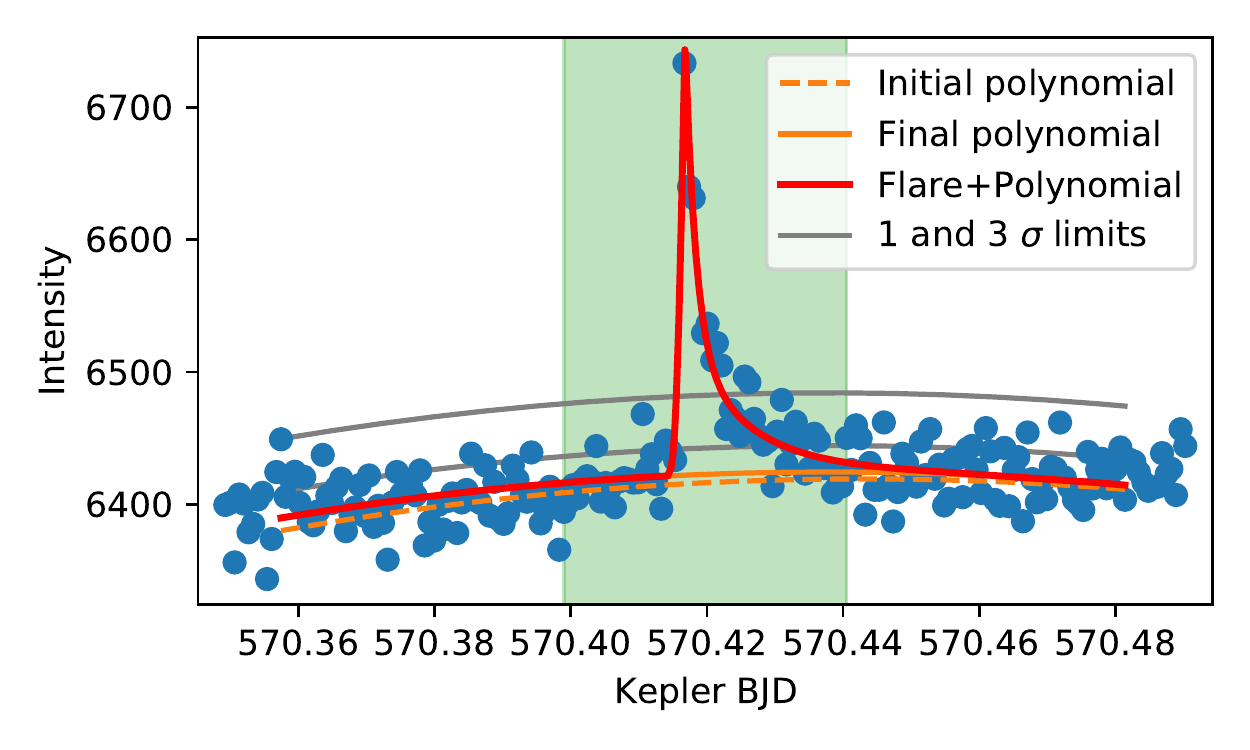}
    \caption{An example for the validation. Green area shows the section marked as a probable flare. Additional flare-length windows before and after the marked region are selected, and fitted by a low-order polynomial. The fit is subtracted from the light curve, and the residual is fitted by another low-order polynomial and an analytical flare model. If the peak of the fit is above the $3\sigma$ noise level, the event is marked as validated. }
    \label{fig:flarefit}
\end{figure}
\begin{figure}
    \centering
    \includegraphics[width=0.5\textwidth]{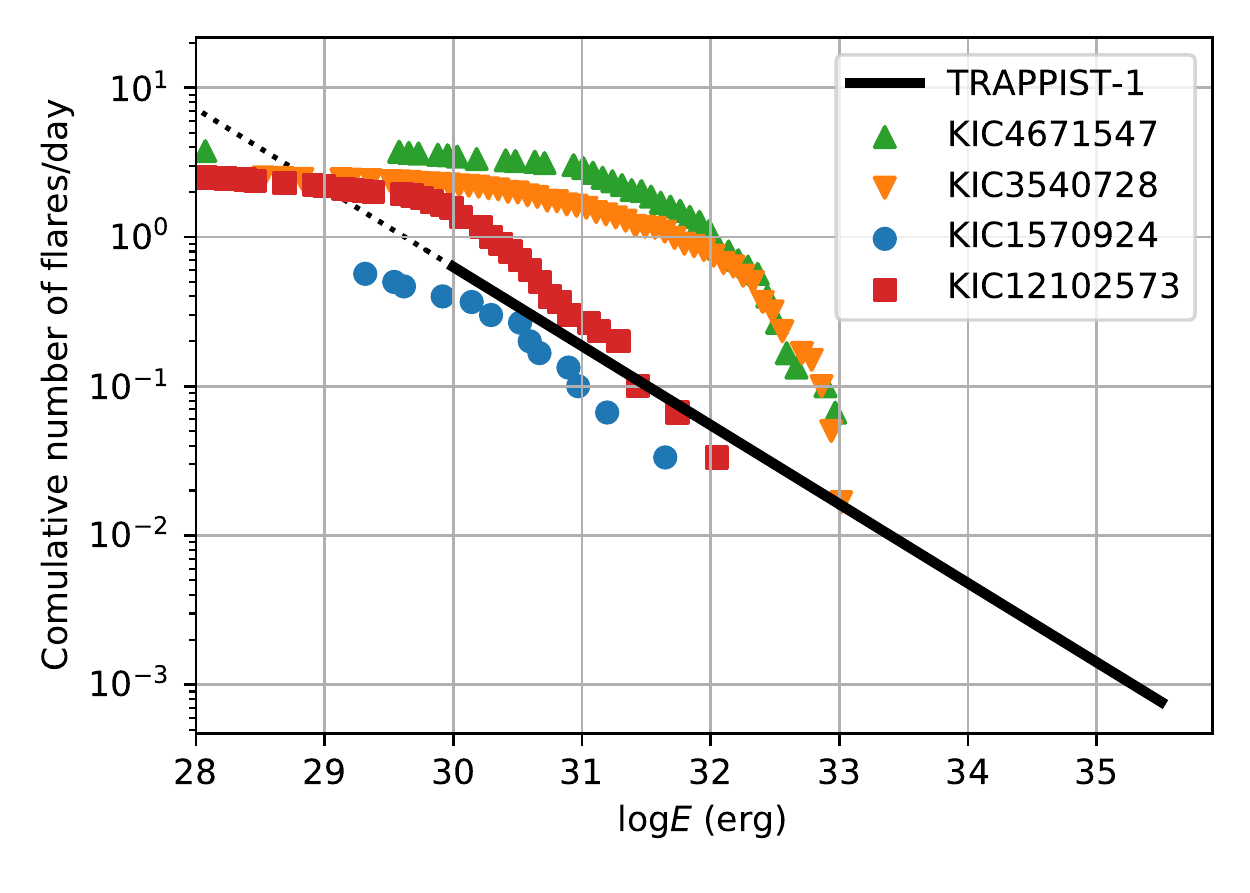}
    \caption{
            Four example flare frequency distributions of KIC 4671547, KIC 3540728, KIC 1570934 and KIC 12102573 (colored symbols). As a comparison we adopted and plotted the fit of FFD of the well-known star TRAPPIST-1 (black line) from \cite{trappist1}. 
            }
    \label{fig:FFD}
\end{figure}

The raw output of the network sometimes contained single false positive points and also smaller sections of flares that were not marked (false negatives), as seen e.g. in Fig. \ref{fig:example-lc}. These issues can be solved by smoothing the raw output by a median filter. The size of the filter, however, should be chosen carefully: a too short kernel size might be not efficient enough, while a too large kernel can filter out shorter events. Fig. \ref{fig:pre-recall} shows the test showing the effect of different kernel size on the precision/recall metrics. 
A window size of 47 and 63 points yielded the best results with a cutoff of 0.2. For the Kepler data we chose a 47 point-wide kernel, as it gave somewhat steadier output, while for the injected flares we used a 63 point filter after averaging the predictions of different K-Fold weights {with a cutoff of 0.3} (see Sect. \ref{sect:performance}).
When selecting the filtering method before validation we favored those options that yielded somewhat higher recall at the cost of losing some precision. Picking a kernel that filters out smaller events among false positives would give higher precision values, but these can be selected later at the validation phase.

The network managed to find the flare events correctly, but in several cases it seemed to neglect part of the eruption. 
To correct these, and to filter out possible remaining false positive detections, we decided to add a further post-processing step to mark these missing points, to detect possible false positives, and also to have an output that is more useful for physical analysis. 

For each marked section, we took a part of the light curve with a length that equals three times the duration of the originally marked points, centered on the middle of the selected window (see Fig. \ref{fig:flarefit}. for an example). 
We fitted these points iteratively with a second-order polynomial filtering points out above 0.5$\sigma$, then subtracted this trend. The residual is used to locate the flare event in the window. To eliminate outliers which can mislead the algorithm, we used a median filter with a width of 3 points. After this, the original light curve was fitted again with a combination of a second-order polynomial and the analytic flare model of \cite{appaloosa}. If the time scale of the light curve variability is comparable to the length of the selected window, the second-order polynomial may not fit properly the background, thus a higher order is needed. To overcome this issue, we repeated the fitting procedure using a third-order polynomial and calculated the Bayesian Information Criterion (BIC; \citealt{BIC}) for both to select the better model. The BIC is defined as
\begin{equation}
    \mathrm{BIC} = n \ln\bigg( \frac{\sum(y_i-\hat{y}_i)^2}{n} \bigg) + k \ln(n),
\end{equation}
where $n$ is the number of points, $k$ is the degree of freedom, and $y$ and $\hat{y}$ are the measured and modeled flux values, respectively. The better the model the lower the BIC. The final model was removed from the light curve and the residual $\sigma$ was used as the noise level to calculate the signal-to-noise ratio. If there still was at least three consecutive points above $1.5\sigma$ noise level an additional fit is performed for complex events. This step was repeated until the highest point went down the noise level.

From the marked events only those can be considered as real flares with high probability that have their peak above $5\sigma$ over the noise level. Below this, the rate of false positive events is increasing rapidly (see Sect. \ref{sect:performance}).
The output also includes the beginning and end time of the events (those points where the fitted model reaches $1\sigma$ level) and the equivalent duration of the flares (their integrated intensity over the event duration). Following the work of \citet{kovari07}, the latter can be used to estimate the flare energy from the quiescent flux of the star assuming a simple black body model, which requires the knowledge of stellar effective temperate and radius. We collected these parameters from the Kepler Input Catalog \citep{KIC-calib}  and calculated the flare frequency distributions (FFD) for some example stars that are shown in Fig. \ref{fig:FFD}.
{The energy distribution characteristics of the recovered flare energies is similar to other methods, but a large number of low-energy events are recovered, which will be helpful in order to obtain information on a wide energy range.  }

\section{Performance}
\label{sect:performance}

\begin{figure*}[p]
    \centering
    \includegraphics[width=0.46\textwidth]{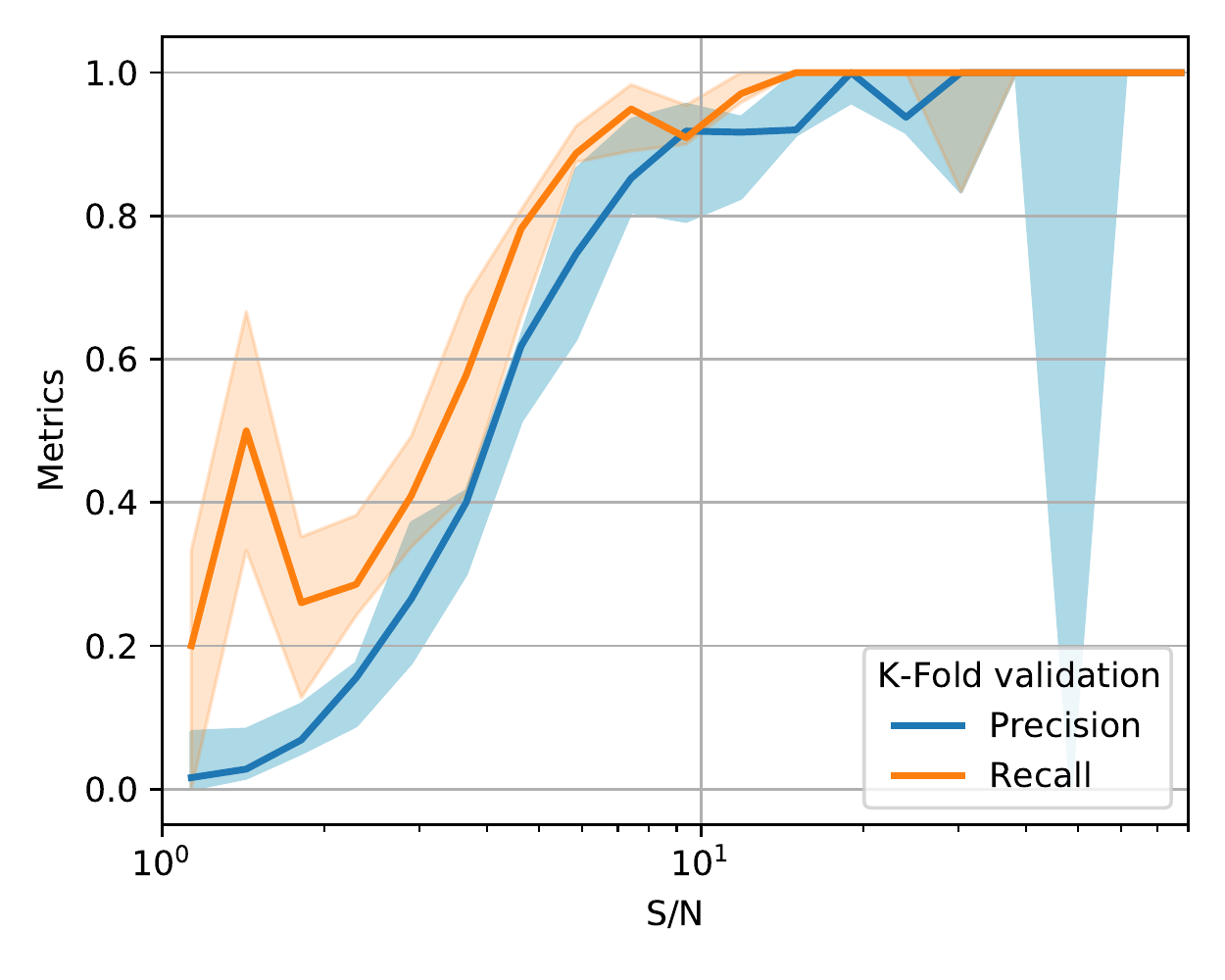}%
    \includegraphics[width=0.46\textwidth]{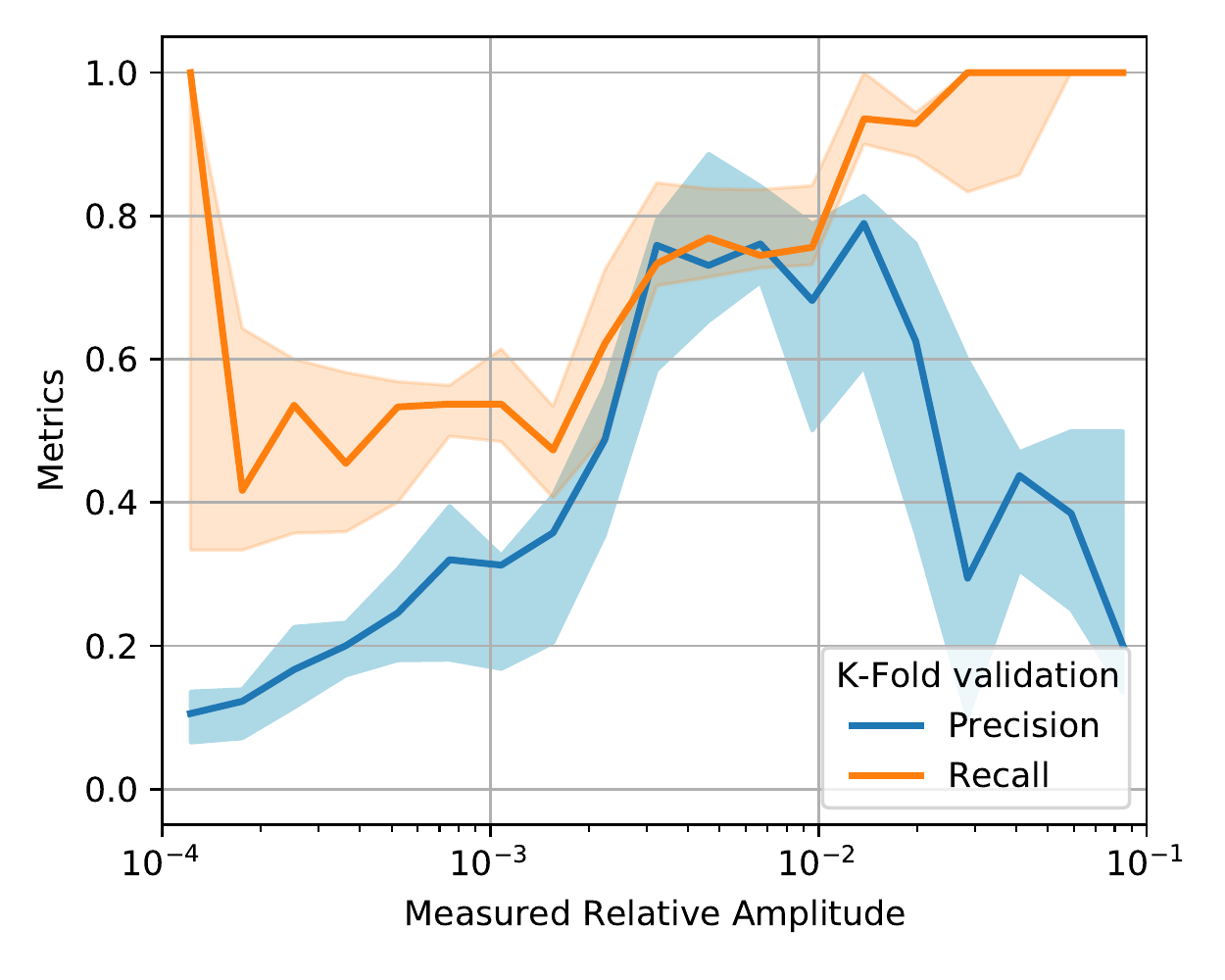}
    \caption{The result of the K-fold cross-validation on the independent test data set. The curves show the precision (blue) and recall (orange) as a function of flare signal-to-noise ratio (left) and relative amplitude (right). The raw predictions were median filtered with a window length of 47. Points are considered to be flares above probability 0.2. Above the $3-5\sigma$ signal level the uncertainty of the models as a result of different training sets is reduced to a few percent, suggesting that the trained model is robust, and that the training sets are sufficiently large. Note the logarithmic abscissa. The low precision in the right plot is caused by the very low number of events.}
    \label{fig:kfold}

    \includegraphics[width=0.46\textwidth]{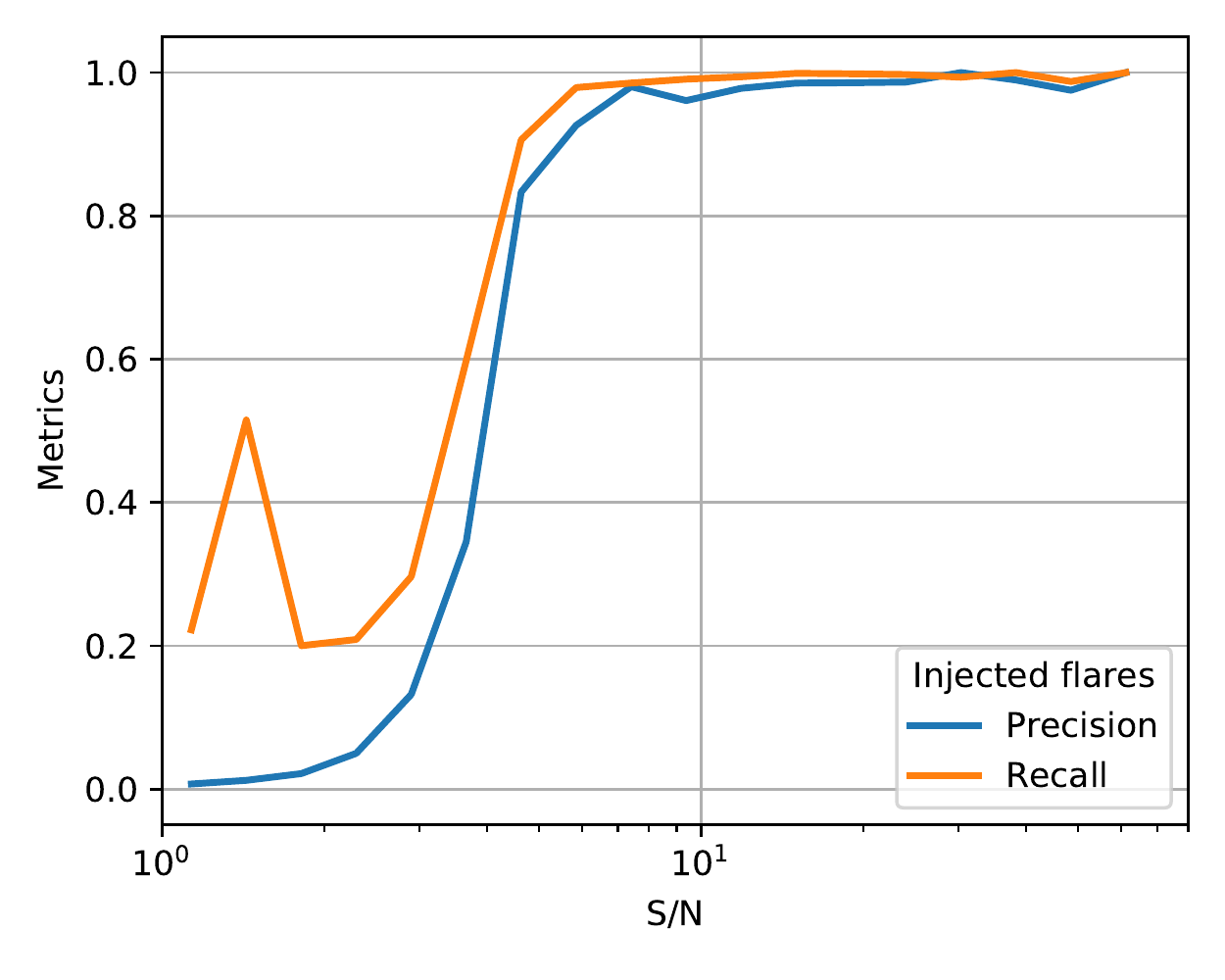}%
    \includegraphics[width=0.46\textwidth]{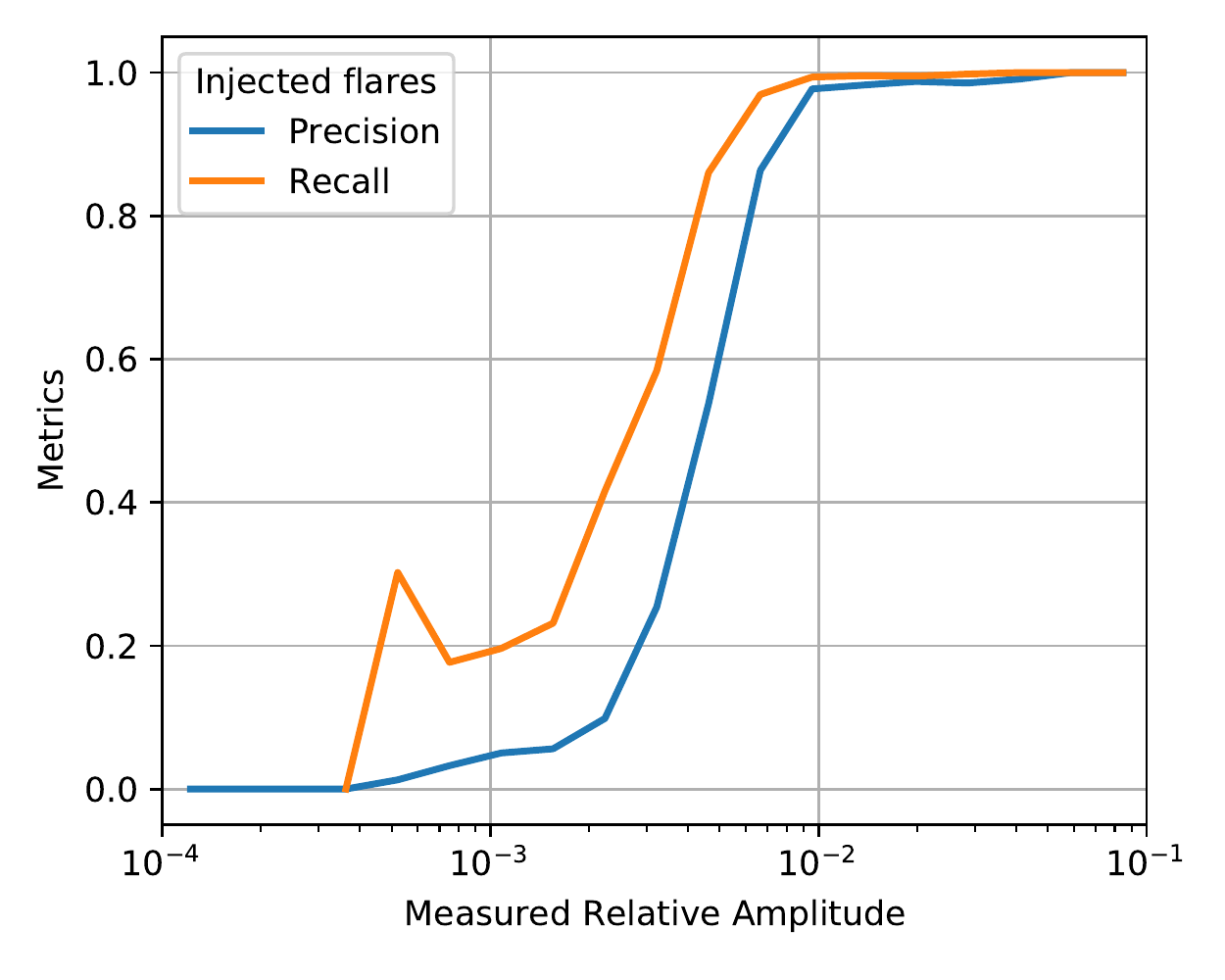}
    \caption{Injection-recovery test using the best performing network, combining the weights of all the K-Fold results. The combined raw predictions were post-processed in the same way as in case Fig.~\ref{fig:kfold}. Note the logarithmic abscissa.
   }
    \label{fig:injected_performance}

    \includegraphics[width=0.46\textwidth]{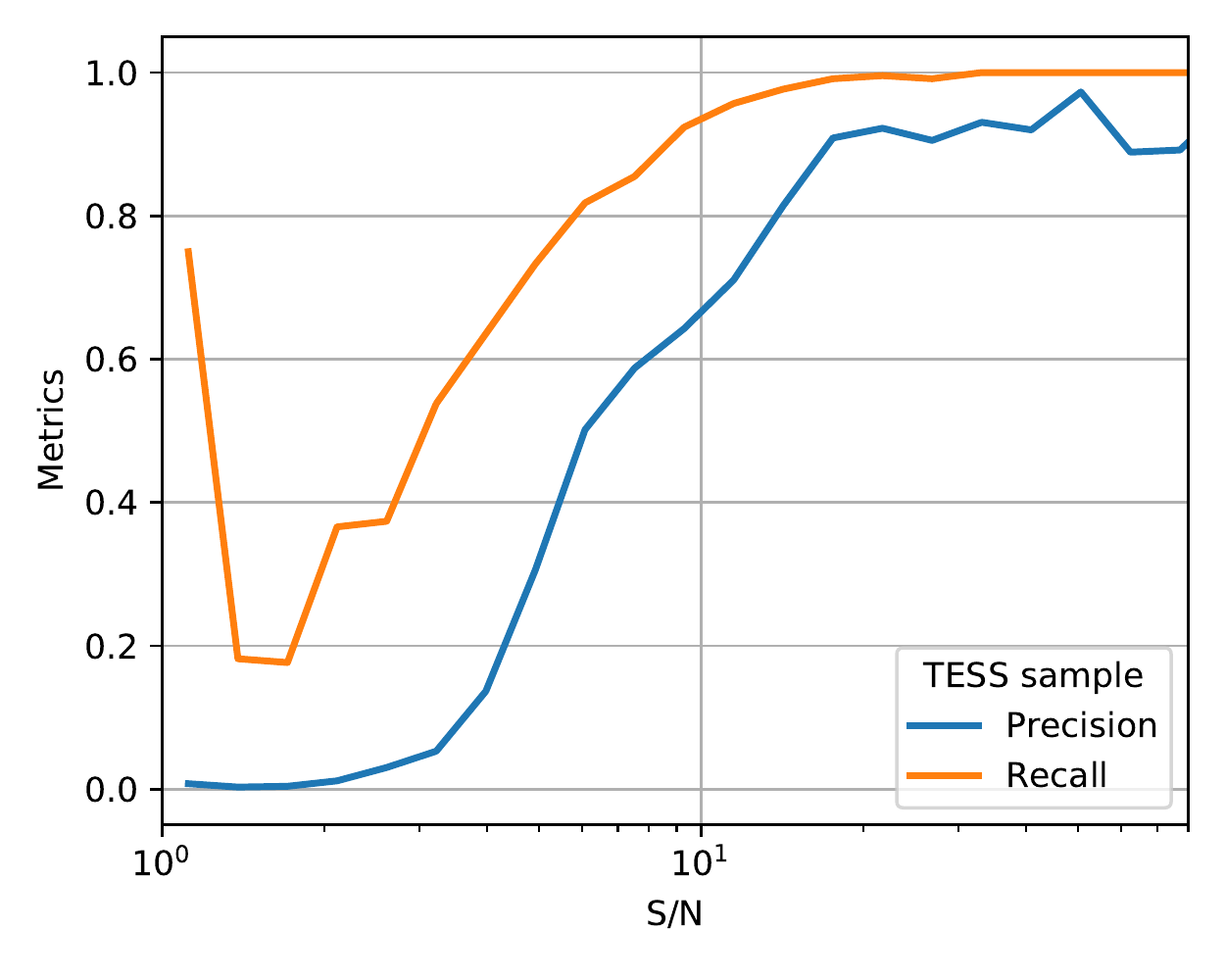}%
    \includegraphics[width=0.46\textwidth]{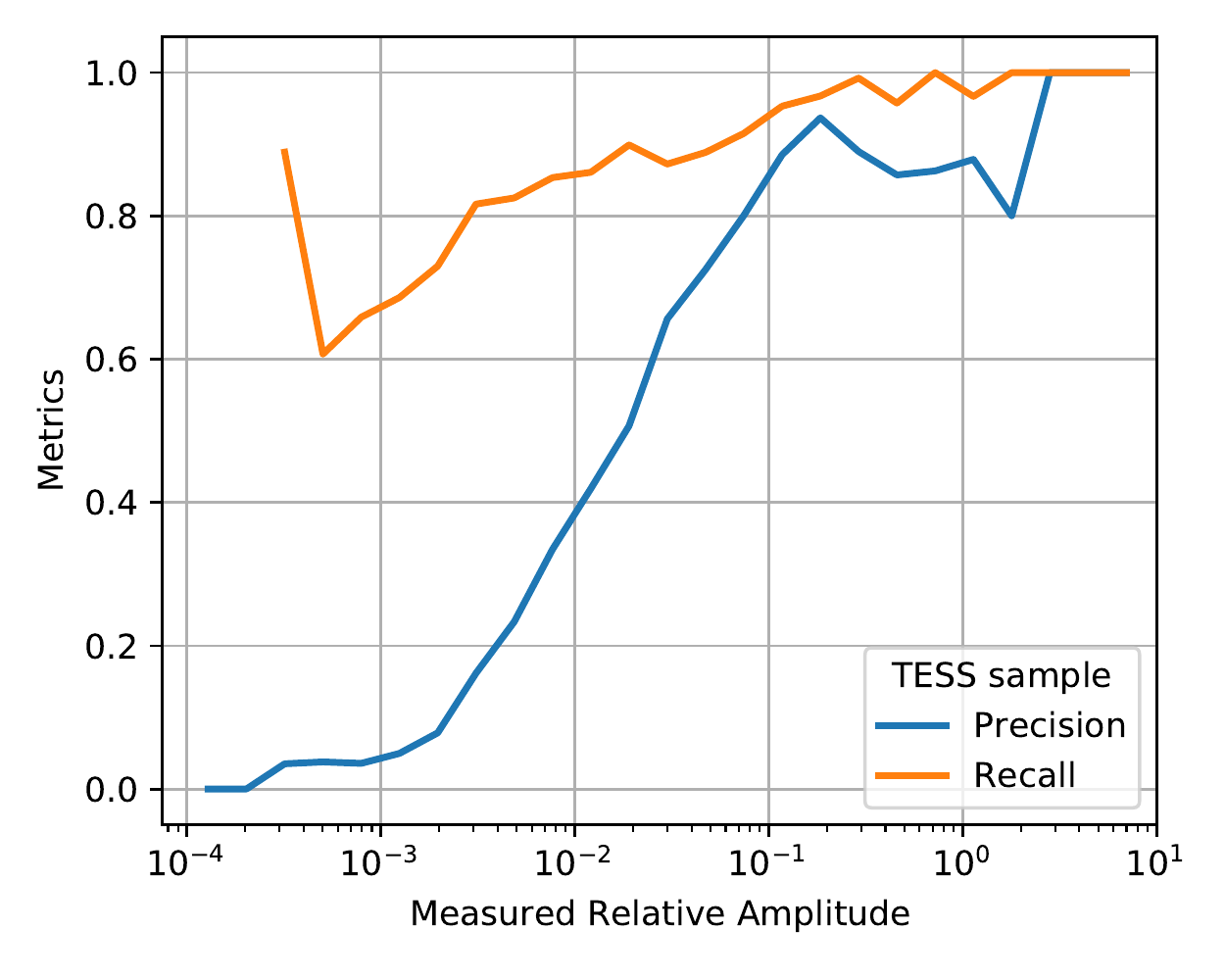}
    \caption{The result of predicting flares in the TESS sample of \cite{TESSflares}. These results were reached using the combined weights of different train runs on the Kepler sample. The raw predictions were median filtered with a window length of 31. Points are considered to be flares above probability 0.5. Note the logarithmic abscissa.}
    \label{fig:tess-test}
\end{figure*}

\begin{figure}
    \centering
    \includegraphics[width=0.46\textwidth]{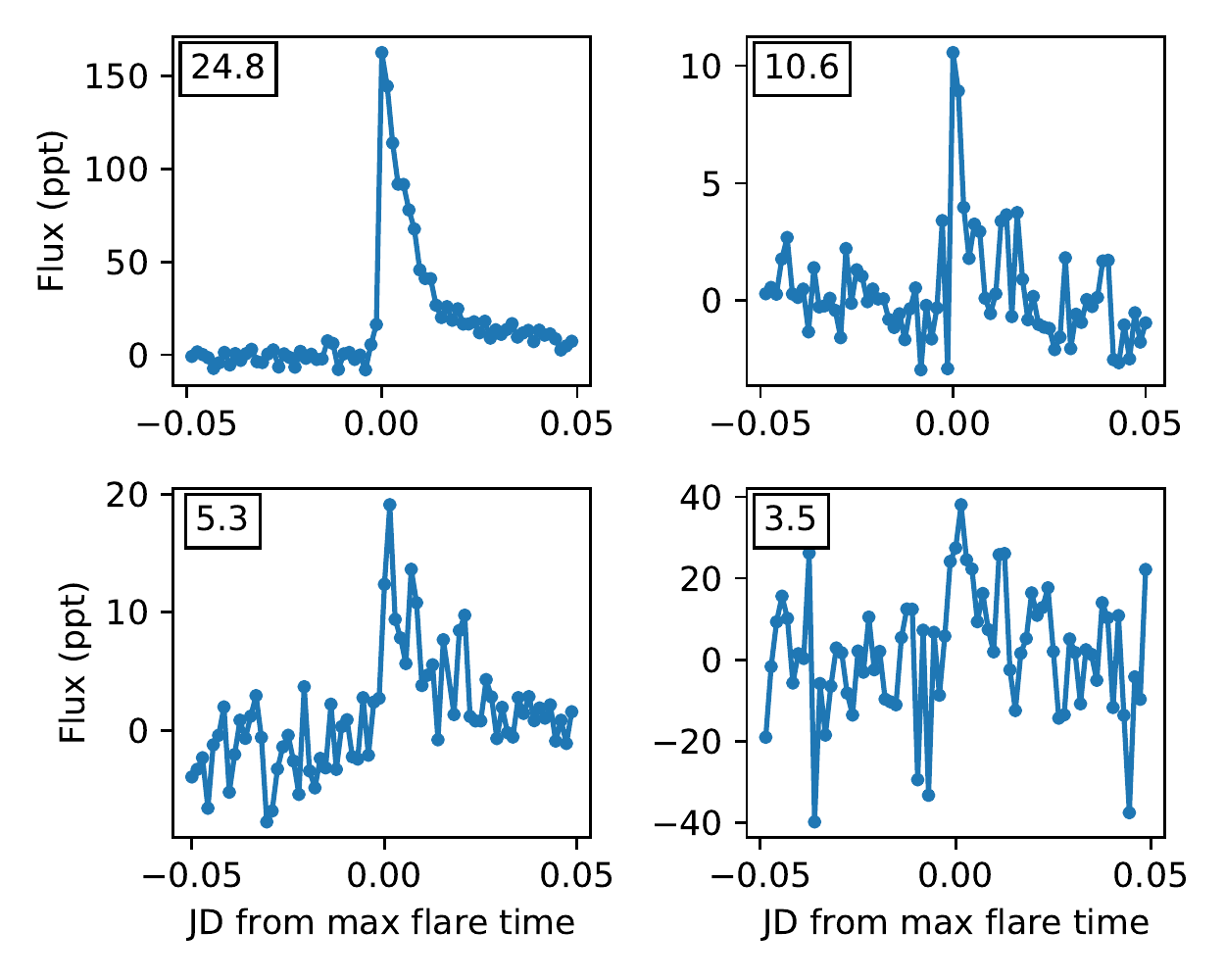}
    \caption{Example high signal-to-noise ratio false positive flares found by our LSTM algorithm in the TESS sample of \cite{TESSflares}. The numbers in the upper left corners show the S/N ratios. Less significant detections are dominated by non-flare events, i.e. genuine false positives.}
    \label{fig:tess-FP}
\end{figure}

We tested the performance of the network on two data sets: a set of Kepler light curves that were not used for training and validation steps,
and light curves with injected flares. Both methods have their drawbacks: with real light curves we can compare the network performance to a human performing the same task, but we do not know the ground truth, i.e., where the flares occurred in reality, and also it is much harder to obtain manually flagged data as generate light curves. Artificial light curves are easy to generate, and we know precisely, where the flares are, but their characteristics might be somewhat different as real events. 

To generate the flare injected data, we selected ten stars from our sample, and removed the flares by smoothing, using multiple passes of a median filter. Then, we injected ten flares into each data set with peaks ranging between 2--15 times the noise level. Those points were marked as part of the event that were above the noise level of the light curve. These steps were repeated 100 times for each star, yielding $10\,000$ flares in $\approx 30\,000$ days of data. 

The performance of the network for the Kepler data set is shown in Fig. \ref{fig:kfold}, where we plotted the precision and recall metrics as a function of the signal-to-noise ratio (left panel) and the measured relative amplitude of flares (right panel). The test shows that above $5\sigma$ signal level 
the LSTM was able to recover $\gtrsim80\%$ of the manually marked flares with a precision of $\gtrsim70\%$. The shape of the metric curves is different in the right plot. The low precision is caused by the very low number of events: above $10^{-2}$ relative amplitude there were only a few dozen events.

For the injected data set, first, we tried prediction with the K-Fold weights. 
We found that, in contrast with the Kepler data set, using only a single K-Fold weight yielded a significantly worse performance. 
Therefore, we tried to weighted average the predictions yielded by different K-Fold weights, which improved the results considerably.
In this case, if one of the networks predicted a point to be a flare with a probability of $>0.98$, we kept it as a flare event, regardless of the results of the other networks.
The results of this test are shown in Fig. \ref{fig:injected_performance}.

The typical runtime for one light curve with a length of one quarter was $\approx 25$ seconds on a GPU with a batch number of $b=16$.

As it can be seen the LSTM is performing better in finding light curves in less noisy areas than finding large amplitude flares in the noise. The non-zero recall number in case of the lowest S/N values is introduced by the flare fitting algorithm. First, it tries to find and fit the significant outlier points in a given flare region regardless of their origin, second, sometimes in the residual of a non-perfect fit, which is caused by the difference between the shape of the model and observed flare, some information remains that is considered to be a flare. Moreover, the number of flagged points within a flare is usually different in the manually and automatically labeled data set, i.e. the duration of flares are different, which gives the opportunity to find different number of flare-like events within a region of a given flare.

\subsection{Evaluation on TESS sample}
\label{sect:tesseval}
We tested the network also on TESS data, in a similar way as it was performed by \cite{CNN-code}.
Light curves were downloaded from MAST using the
 \verb+Lightkurve+ 
tool\footnote{\url{https://docs.lightkurve.org}} \citep{lightkurve}.
Labels for flares were obtained from \cite{TESSflares}. 
This set contains a large amount of flaring light curves, but the labels are not perfect, there is a number of missing flares, or incorrect labels: these events will bias our comparison yielding incorrectly false positive or false negative detections. As the RNN requires equidistant sampling, the light curves were resampled to 2 minute cadence, which is equal to the TESS short-cadence sampling.

We predicted flares using all the weights saved during the Kepler K-fold test, and additionally we trained the network with all available Kepler light curves and utilized those weights as well. In the individual tests we were not able to recover as many flares as in case of the Kepler data. As it turned out, the main problem was the normalization of light curves. If we transform the flux values to the the range of 0--1, the large amplitude flares suppress the small events making those impossible to identify. To overcome this problem, we sigma clipped the large flares before normalization. However this step raised another issue, the height of high amplitude flares went way above unity, which made the RNN uncertain, yielding much lower probabilities for these previously well recovered events. Thus, we somehow had to combine the results of different predictions. After an extensive test, we found that the best solution is to use three weights rather than only one, predict the location of flares in light curves normalized by both methods and weighted average their predictions. The weights are unity, unless a given instance predicts an event with a probability larger than 0.9, then its prediction takes over the others. This way both small and large events are captured, and the false negative ratio is minimized.

The precision and recall values as a function of signal-to-noise ratio and relative flare amplitude are shown in Fig. \ref{fig:tess-test}. After a precision--recall test, similar to the one presented in Fig. \ref{fig:pre-recall}, the raw predictions were median filtered with a window length of 32 and points were considered to be flares above probability threshold of 0.5.
Although, the TESS data have different sampling and noise characteristics that may influence the results, our test showed that we are able to recover a similar number of flares as in case the Kepler data shown in Fig. \ref{fig:kfold}--\ref{fig:injected_performance}. The shape of the precision curves are similar too, but the absolute vales are 10--40\% lower. Plotting the individual events marked as flares by our network, we found flares in the order of 4--5000 missing from the original catalog with similar characteristic to those of plotted in Fig. \ref{fig:tess-FP}. Considering the number of flares in the catalog of \cite{TESSflares}, which is about 8600, this difference is understandable. 
This test showed that even the current network is able to generalize and find flares -- with similar effectiveness -- in completely new data having previously unseen sampling and characteristics.

The \texttt{flatwrm2} code and the network used by \cite{CNN-code} use both different data sets and approaches. The latter is trained on TESS data that have different characteristics, and has a larger  flare sample compared to the Kepler sample. 
Another important difference lies in the scaling and normalization of the input data: while we normalize the light curve as a whole, in case of \texttt{stella} each window is normalized separately. Therefore, \texttt{flatwrm2} might perform worse in cases where the light curve itself is peculiar, and the normalization process yields data with too large or too small events that are not recognized by the neural net. On the other hand, \texttt{flatwrm2} could outperform \texttt{stella} in the case of atypical flare shapes, complex flares or long flares, due to the network remembering its previous states thanks to the forget/update gates in LSTM. While \texttt{flatwrm2} was trained on Kepler short-cadence data only (and artificial light curves based on Kepler data), the LSTM network could be better in generalization. However, a qualitative comparison of the two methods is currently not possible given their mission-oriented training sets.

\section{Failures}
\label{sect:failures}

Research is rarely a linear process with forming the idea, working through it and concluding the results. This is especially true for developing machine learning methods -- in this section we will shortly summarize those experiments, that did not work at all or gave inferior results.

Initially we tried selecting the best candidates based on simple model data only, and use this weights on real data, but we found that over-simplified light curve models yield unusable model weights and architectures.

As mentioned in Sect. \ref{sect:input}, standardizing input data is crucial. Here, simply dividing with the mean or median of each data set did not work, and also normalizing the data with their standard deviation proved problematic: in some cases larger flares were scaled up, and they got neglected by the network.

We experimented with a large number of network architectures that did not converge at all, or did not yield the expected results. We tried:
\begin{itemize}
\item simple dense networks with different sizes;
\item convolutional networks;
\item hybrid networks utilizing both recurrent nets and dense or convolutional networks in parallel
\end{itemize}
with various inferior results (our takeaway lesson was that a more complex network is not necessarily performing better).
Adding an extra dense layer with 32 units after the selected recurrent layers did also not improve the results.
Bidirectional recurrent networks (that analyze the input data in both direction) worked in some cases slightly better, but at the cost of doubling the runtime, therefore we decided to use one-directional analysis (with this the prediction time for a single-quarter Kepler light curve is $\approx13$ seconds on GPU with a batch number of $b=64$).

Changing the window size for training did not improved the results either. Using a shorter window, the number of false positive detections increased significantly, while larger windows tend to smear out smaller events.

\section{Code availability}
\label{sect:availability}

To make our recurrent neural network a useful tool for finding flares in the Big Data era, we prepared a user-friendly code, dubbed as \texttt{flatwrm2}, which is available through a GitHub repository\footnote{\url{https://github.com/vidakris/flatwrm2}}, and can be installed via PyPI. 
Passing a flux timeseries, 
\texttt{flatwrm2} 
will prepare the light curve, predict and validate the possible flare events. The user will encounter a bunch of output parameters, such as the location, FWHM, relative amplitude and signal-to-noise ratio of flares, of which the latter can be used to select the more likely intrinsic events. Besides the raw code, we have written a tutorial Notebook which guides the user through the prediction and validation steps.

\section{Summary}
\label{sect:summary}

We presented an experiment to detect flares in space-borne photometric data using deep neural networks (NNs). After testing a large number of network architectures, we concluded that the best performing network was a recurrent neural network based on using Long Short-Term Memory (LSTM) layers. After training a network using a set of artificial light curves, and real short-cadence Kepler observations of flaring and non-flaring stars, the network was not only able to detect flares with peaks over $5\sigma$ with 80--90\% recall and precision, but also was capable to distinguish false signals that typically confuse flare-finding algorithms (e.g. maxima of RR Lyr stars) from real flares. 

In the future we plan to analyze a large number of Kepler and TESS data to gain a wider knowledge of flaring stars. The code with the weight files are available in the public domain for the scientific community.

\section{Software}
Python \citep{Python3}
--Numpy \citep{numpy}
--Pandas \citep{pandas}
--Scikit-learn \citep{scikit}
--Tensorflow \citep{Tensorflow}
--Keras \citep{Keras} 
-- Matplotlib \citep{matplotlib} 
--scipy \citep{scipy}
--PyMacula \citep{macula}
--Lightkurve \citep{lightkurve}

\begin{acknowledgements}
The authors would like to thank deeplearning.ai and Coursera for hosting their online course on different aspects on machine learning, without which this work would never be possible.
{We thank the referee for their swift replies and the valuable remarks that considerably helped us to improved the manuscript and the code availability.} 

   BS was supported by the \'UNKP-19-3 New National Excellence Program of the Ministry for Innovation and Technology.

This project has been supported by the Lend\"ulet Program  of the Hungarian Academy of Sciences, project No. LP2018-7/2020, the 
NKFI KH-130526, 
NKFI K-131508 and 
2019-2.1.11-TÉT-2019-00056
grants. 

On behalf of \textit{"Analysis of space-borne photometric data"} project we thank for the usage of ELKH Cloud (\url{https://science-cloud.hu}) that helped us achieving the results published in this paper.

Authors acknowledge the financial support  of the Austrian-Hungarian  Action  Foundation (95 \"ou3, 98\"ou5, 101\"ou13). 

\end{acknowledgements}

%
%

\bibliographystyle{aa} 
\bibliography{references}

\begin{appendix} 

\section{Performance of different architectures}
\label{sect:appendix}

\begin{figure*}
\includegraphics[width=\textwidth]{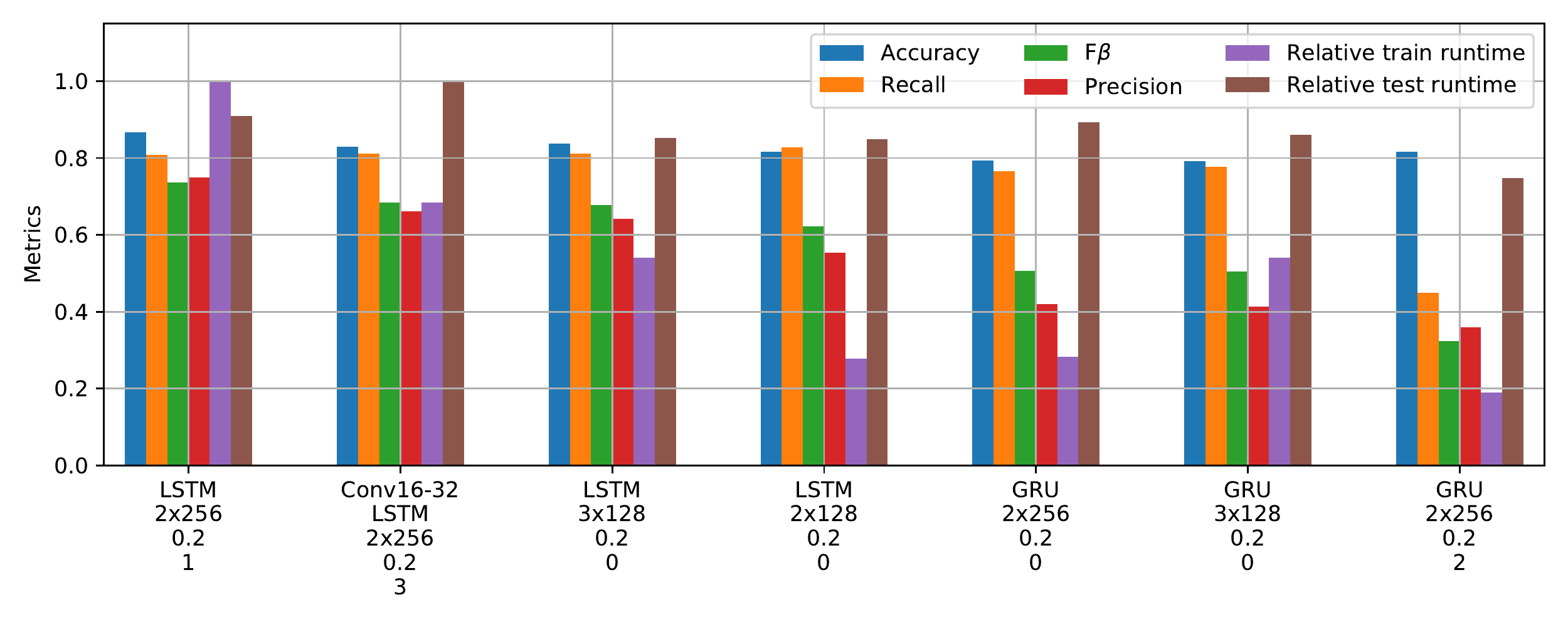}
\caption{The performance of the designed and tested RNN architectures. The bars show the accuracy (blue), recall (orange), F$\beta$ (green), precision (red) and the relative runtimes of training (pink) and evaluating (brown) compared to the slowest instance, respectively. The networks that loss started to increase dramatically after some epoch were removed.}
\label{fig:compare_RNNs}
\end{figure*}

In Fig. \ref{fig:compare_RNNs}, we show the performance of the RNN architectures we designed, trained and tested on the same data set. As a single metric is not enough to select the best one, we plotted the accuracy, recall, precision, F$\beta$, along with the training and evaluation time compared to the slowest instance, respectively. The TP, FP and FN values were calculated from the maxima times of detected flares. The accuracy is biased as the number of TN flares, considering all the points the are not flagged, are orders of magnitude larger than the flagged points. The rows in labels are the name, units and layers of the kernels, the dropout value and number of the trained dense layers after the RNN units. The last dense layer that is used to link the outputs is not shown. In case of the second network, we used two CNN layers with filter sizes of 16 and 32 before the RNN. We selected the third architecture to work with, as it yielded almost the same results as the first two, but with a reduced training time of only 40\% and 20\%.

\end{appendix}

\end{document}